\documentclass[aps,prl,twocolumn,longbibliography]{revtex4-1}
\usepackage[utf8]{inputenc}
\usepackage[pdftex]{graphicx}
\usepackage{amsmath}
\usepackage[amssymb,Gray]{SIunits}

\begin{document}


\title{Comment on ``Do Gedankenexperiments compel quantization of gravity''}


%
\author{Andr\'e Gro{\ss}ardt}
\email[]{andre.grossardt@uni-jena.de}
\affiliation{Institute for Theoretical Physics, Friedrich Schiller University Jena, Fr\"obelstieg 1, 07743 Jena, Germany}


\date{\today}

\begin{abstract}
There is no conclusive theoretical argument that would require the quantization of the gravitational field. A paradox presented by Mari et al.\ [Sci. Rep. 6, 22777 (2016)] and its resolution by Belenchia et al.\ [Phys. Rev. D 98, 126009 (2018)] did not change anything about this assertion. Despite being interesting in other respects, efforts of a recent work by Ryding, Aurell, and Pikovski [arXiv:2107.07514] to further establish the inconclusiveness of the question of necessity of quantization are futile.
\end{abstract}


\maketitle



In a recent article, Rydving et al.~\cite{rydvingGedankenexperimentsCompelQuantization2021} discuss a paradox~\cite{mariExperimentsTestingMacroscopic2016} that acquisition of which-path information for an interference experiment through the gravitational field could be exploited for superluminal signalling. Belenchia et al.~\cite{belenchiaQuantumSuperpositionMassive2018} previously explained how a consistent description, including decoherence from graviton emission as well as vacuum fluctuations and, thereby, treating the gravitational field as quantized, avoids this paradox.

Rydving et al.\ propose a new resolution, implying only the Planck length $l_P$ as a minimum resolution. The bound $\Delta x > l_P$ for Bob's particle then becomes a postulate rather than a consequence of vacuum fluctuations~\cite{belenchiaQuantumSuperpositionMassive2018}, and instead of the condition that Alice's interference experiment must be slow enough to not emit a single graviton one poses the stricter condition that interference fringes must be a Planck length apart.

I would like to comment on three aspects of their work, with a particular focus on the final one about the meaning of the discussed paradox with regard to semiclassical gravity, where I believe that the authors are taken in by a common misconception about the implications of the work of Belenchia and others.

\paragraph{Necessity of the octopole moment}
Firstly, Rydving et al.\ make the important observation that in the symmetric set-up considered by Belenchia et al.\ the relevant lowest order gravitational force on Bob's particle stems from the octopole moment of Alice's particle. It has been previously discussed~\cite{belenchiaQuantumSuperpositionMassive2018} that there is no contribution to the dipole moment as the entire system of particle plus laboratory must be considered whose mass center is conserved. The quadrupole moment, however, depends only on the magnitude of the displacement of Alice's particle; dependence on direction only occurs for the octopole~\cite{rydvingGedankenexperimentsCompelQuantization2021}.

Although this is an interesting realization, it is of no further consequence for the arguments in either work~\cite{belenchiaQuantumSuperpositionMassive2018,rydvingGedankenexperimentsCompelQuantization2021}. It can also be easily mitigated by chosing an asymmetric path for Alice's particle, e.\,g.\ in a Mach-Zehnder interferometer, where the shift with respect to the intial/final rest frame of the laboratory is 0 or $d$ rather than $\pm d$.

\paragraph{Universality of the Planck length restriction}
Secondly, Rydving et al.\ dispose of the necessity of quantized radiation by instead requesting that the Planck length can ``be assumed as a lower cutoff beyond which no conclusions can be drawn''~\cite{rydvingGedankenexperimentsCompelQuantization2021}. The difference in trajectory for Bob's particle can then---for whatever reason---only be resolved if it is larger than $l_P$. For Alice's particle, the requirement that interference fringes are at least a Planck length apart results in the mass bound $m_A < T_A/d$ which implies a limit on the quadrupole $\mathcal{Q}_A < d T_A$ stricter than the limit $\mathcal{Q}_A < T_A^2$ given in~\cite{belenchiaQuantumSuperpositionMassive2018}. The conclusion that this bound prevents any superluminal exchange of information remains valid, of course.

The ad hoc restriction of sub-Planckian spatial resolutions may, however, not be as universal as the restrictions discussed in~\cite{belenchiaQuantumSuperpositionMassive2018}. For a system of $N$ identical particles, all localized with an uncertainty $\Delta x > l_P$, the center of mass is localized with uncertainty $\Delta x / \sqrt{N}$ which can fall below the Planck length. If Bob's particle is a crystal, for instance, a force that results in a displacement of only a few atoms could well result in observable effects, e.\,g.\ in diffraction, while displacing its mass center by less than a Planck length.
Whether or not this is possible depends on the details of the gravitational interaction and the exact nature of the Planck length resolution limit.

\paragraph{Implications for semiclassical gravity}
Finally, the conclusion drawn by Rydving et al.\ is that ``the need to quantize gravity as a quantum field theory remains inconclusive''~\cite{rydvingGedankenexperimentsCompelQuantization2021}. I agree but assert that this inconclusiveness has never been in question. Although one might misinterpret Belenchia et al.~\cite{belenchiaQuantumSuperpositionMassive2018} in this manner,  their paper makes no such claim.

The paradoxical situation~\cite{mariExperimentsTestingMacroscopic2016} occurs when combining quantum mechanics with an action at a distance---regardless whether gravitational or electromagnetic in its nature. Belenchia et al.\ show how the paradox \emph{can} be resolved by treating the interaction properly within the framework of quantum field theory, i.\,e. the paradox does \emph{not} point towards an inconsistency of quantized electrodynamics or gravity.

It does not imply that quantized gravity is the \emph{only} theory resolving the paradox. In fact, the paradox disappears in a quite obvious way in the most common semiclassical model~\cite{mollerTheoriesRelativistesGravitation1962,rosenfeldQuantizationFields1963} where gravity is sourced by the expectation value of stress-energy. In this model, Alice's superposition always generates the same gravitational potential depending on both paths. The trajectory of Bob's particle, therefore, will always be consistent with the average quadrupole $\mathcal{Q}_A \sim d^2$ and the average octopole $\mathcal{O}_A = 0$. Which-path information can never be transferred.

In a more general version, Bob's inability to extract which-path information is directly related to the impossibility to create entanglement via a semiclassical gravitational link, which resolves the paradox for any semiclassical model to be tested for in similar experimental proposals~\cite{boseSpinEntanglementWitness2017,marlettoGravitationallyInducedEntanglement2017}.

What is actually shown by Rydving et al.~\cite{rydvingGedankenexperimentsCompelQuantization2021} is that even a model that does allow for entanglement may not require quantization of radiation in the form of gravitons. Instead, a minimal spatial resolution may be sufficient to resolve the paradox.
Regarding the quantization of gravity, Gedankenexperiments \emph{do not} compel it \emph{and never have}. What Rosenfeld so wisely assessed more than half a century ago remains true: ``Even the legendary Chicago machine cannot deliver the sausages if it is not supplied with hogs.''\cite{rosenfeldQuantizationFields1963}

\paragraph{Acknowledgment}
\begin{acknowledgments}
I gratefully acknowledge funding by the Volkswagen Foundation.
\end{acknowledgments}

%

\end{document}